\documentclass[twocolumn,showpacs,preprintnumbers,amsmath,amssymb,floatfix]{revtex4}

\usepackage{graphics}
\usepackage{color}
\usepackage{psfrag}

\begin{document}

\title{Oscillating elastic defects: competition and frustration}

\author{
J.~Barr{\'e}, A.~R.~Bishop, T.~Lookman, A.~Saxena}

\affiliation{Theoretical Division, Los Alamos National Laboratory, Los
  Alamos, NM 87545, USA}

\date{\today}

\begin{abstract}  
  
  We consider a dynamical generalization of the Eshelby problem: the
  strain profile due to an inclusion or ``defect" in an isotropic
  elastic medium. We show that the higher the oscillation frequency of
  the defect, the more localized is the strain field around the
  defect.  We then demonstrate that the qualitative nature of the
  interaction between two defects is strongly dependent on separation,
  frequency and direction, changing from ``ferromagnetic" to
  ``antiferromagnetic" like behavior. We generalize to a finite density
  of defects and show that the interactions in assemblies of defects
  can be mapped to XY spin-like models, and describe implications for
  frustration and frequency-driven pattern transitions.
 
\end{abstract}


\maketitle

In his seminal 1957 paper~\cite{Eshelby57}, Eshelby derived the strain
fields created by an inhomogeneous ellipsoidal inclusion in an
isotropic elastic medium. This result is a cornerstone of the theory
of inhomogeneous elastic media, now routinely used in the physical and
engineering sciences. The result is a statement of how a distortion is
accommodated in the host material and implies that a local
perturbation induces long range strain fields, slowly decaying as
$1/r^d$ in $d$ dimensions ($d\geq 2$).  Eshelby's work considers
static inhomogeneities or strain ``defects" only. However, for many
applications in physics and materials science, we are interested in
the cooperative behavior of inhomogeneities in which the strain is
varying or oscillating in time with a given frequency. We find that
this situation is, as a function of the defect density and oscillation
frequency, inherently frustrated, resulting in self-organization of
the patterns, competing ground states, and sensitivity to internal and
external perturbations. Such ``dynamic" defects arise as small polarons
in directionally-bonded transition metal oxides, including
high-temperature superconductors, colossal magnetoresistance materials
and ferroelectrics~\cite{ref_polarons}. The collective behavior of
these polarons in a crystal undergoing distortions, with their
coupling to charge, spin or polarization, is believed to determine the
overall macroscopic response. Our work also has ramifications for the
non-destructive evaluation of elastic media. Methods in this field are
typically based on the vibrational response from a defect-free
crystal.  Here we characterize the behavior of oscillating defects
(external oscillatory fields inducing specific defect patterns and
responses) extending the conventional analysis to describe the
response of elastic media in the presence of such defects.\\

The dynamical generalization of the Eshelby problem has been
investigated in the context of engineering sciences for spherical
inclusions~\cite{Mikata90}, and very recently for inclusions of
various shapes~\cite{Wang05}. However, such studies have focused on
evaluating displacement fields as solutions to numerical boundary
value problems. Our objective here is to understand the effect of
dynamics on the nature of the elastic interaction itself and its
influence on the collective behavior of assemblies of defects, using
the formalism developed in Ref.~\cite{prb}, which allows for tractable
analytic calculations. We consider the dynamical Eshelby problem for
localized oscillating defects in two dimensions for simplicity, and
show that, although the strain fields still decay as $1/r^2$ far from
the defect, the frequency fundamentally affects the nature of
deformation. As expected, the higher the frequency, the more localized
is the deformation. This renders the interaction between two defects
strongly frequency (and direction) dependent, but the very nature of
the interaction changes from ``ferromagnetic" to ``antiferromagnetic"
like behavior as a function of separation and frequency. We
subsequently generalize our results to a finite density of defects.
This allows us to demonstrate the implications for frequency-driven
patterning transitions and phase locking in assemblies of defects by
mapping the elastic interaction energy between defects into XY
spin-like models with competing interactions.

We use a strain only representation~\cite{prb}. The state of strain is
defined by three fields $e_i$ related to the displacements along the x-axis
($u$) and the y-axis ($v$) as follows: $e_1=(u_x+v_y)/\sqrt{2}$,
$e_2=(u_y+v_x)/\sqrt{2}$, $e_3=(u_x-v_y)/\sqrt{2}$ (the subscripts $x$
and $y$ indicate differentiation). We write an elastic energy which
includes gradient terms:
\begin{eqnarray}
E&=&\int\left[ \frac{A_1}{2} e_1^2+\frac{A}{2} (e_2^2+e_3^2)\right]~d\vec{r} 
\nonumber\\ 
&& +\frac{g}{2}\int \left[ (\nabla e_1)^2+ (\nabla e_2)^2+(\nabla e_3)^2 
\right]~d\vec{r}~,
\label{eq:energy}
\end{eqnarray}
where $A_1$ and $A$ are the bulk and shear moduli of the isotropic
elastic material, respectively, and $g$ is a strain gradient
coefficient. Assuming first an overdamped dynamics (we comment later
on the underdamped case), the equations to be solved read, in real
space:
\begin{eqnarray}
\label{eq:eqtot1}
\dot{e_1} &=& -(A_1 + g \nabla^2)e_1 - \nabla^2 \lambda~, \\
\dot{e_2} &=& -(A + g \nabla^2)e_2 +2 \partial x \partial y\lambda~, \\
\dot{e_3} &=& -(A + g \nabla^2)e_3 +(\partial^2 x-\partial^2 y)\lambda~, \\
0 &=& \nabla^2 e_1-2\partial x \partial y e_2 +(\partial^2 x-
\partial^2 y)e_3 ~.
\label{eq:eqtot4}
\end{eqnarray}
The last equation is the compatibility condition, reflecting the fact
that the three strain fields are not independent; $\lambda$ is a
Lagrange multiplier enforcing this constraint. The boundary conditions
are taken to be periodic in space. We now add $n$ oscillating defects
localized around the positions $(\vec{r_i})_{i=1...n}$. We assume for
specificity that the oscillation is in $e_3$.  Considering strictly
point defects, that is $e_3(\vec{r_i},t)=e_0
\delta(\vec{r}-\vec{r_i})\sin (\omega_0 t +\varphi_i)$, with
$\omega_0$ and $\varphi_i$ denoting oscillation frequency and phase,
would lead to unphysical logarithmic divergences of the Green
functions for the system of
equations~(\ref{eq:eqtot1})-(\ref{eq:eqtot4}).  Thus, we enforce the
regularization: $\int e_1(\vec{r_i},t) g_{\sigma}(\vec{r})~d\vec{r} =
\int e_2(\vec{r_i},t) g_{\sigma}(\vec{r})~d\vec{r} = 0$, $\int
e_3(\vec{r_i},t) g_{\sigma}(\vec{r})~d\vec{r} = e_0 \sin (\omega_0 t
+\varphi_i)$,
where $g_{\sigma}(\vec{r})=\exp(-r^2/2\sigma)/(2\pi\sigma)$, and
$\sigma$ is chosen small enough that the defect is physically
localized. We have checked that our results, which apply to the far
field created by the defects, do not qualitatively depend on the
regularization.

We start with the case of one defect. As the problem is linear, we can
construct the solution as a superposition of $\phi^{(ij)}_{\pm}$, the
elementary solutions of the problem, defined as follows:
\begin{eqnarray}
\dot{\phi}^{(11)}_{\pm} &=& -(A_1 + g \nabla^2)\phi^{(11)}_{\pm} - 
\nabla^2 \lambda +g_{\sigma}(\vec{r}) e^{\pm i\omega_0 t}~,\\
\dot{\phi}^{(12)}_{\pm} &=& -(A + g \nabla^2)\phi^{(12)}_{\pm} 
+2 \partial x \partial y\lambda~,\\
\dot{\phi}^{(13)}_{\pm} &=& -(A + g \nabla^2)\phi^{(13)}_{\pm} 
+(\partial^2 x-\partial^2 y)\lambda~,\\
0 &=& \nabla^2 \phi^{(11)}_{\pm}-2\partial x \partial y\phi^{(12)}_{\pm}  
+(\partial^2 x- \partial^2 y)\phi^{(13)}_{\pm}.
\end{eqnarray}
The $\phi^{(2j)}_{\pm}$ and $\phi^{(3j)}_{\pm}$ are solutions of the
same set of equations with the oscillatory excitation in the second
and third equations, respectively. The $\phi^{(ij)}_{\pm}$ can be
analytically calculated in Fourier space. Writing the $e_i$ as linear
combinations of the $\phi^{(ij)}_{\pm}$, and finding the coefficients
by enforcing the regularized oscillating defect conditions, we obtain the
expressions for the fields. 
We give here explicitly the expression for $\phi^{(12)}_{+}$ (the other 
$\phi^{(ij)}_{\pm}$ have similar characteristics):
\begin{equation}
\phi^{(12)}_{+} = \frac{2k_xk_y}{k^2}
\frac{g_{\sigma}(\vec{k})e^{i \omega_0 t}}{A+A_1+2gk^2 + 2 i\omega_0 t}~.
\label{eq:phi12}
\end{equation}
Two important results can be deduced from this expression. First, due
to the compatibility condition, $\phi^{(12)}_{+}$ is not continuous
around $\vec{k}=0$. This creates the same $1/r^2$ tails as in
the static case. The Eshelby result thus extends to oscillating
defects. Second, from Eq.~(\ref{eq:phi12}) we see that the larger
$\omega_0$, the more spread out is $\phi^{(12)}_{+}(\vec{k})$. In real
space this implies that the larger $\omega_0$, the more localized is
the deformation created by the defect. This will be of primary
importance for the interactions between defects, as detailed below.\\

\begin{figure}
\resizebox{0.48\textwidth}{.24\textheight}
{\includegraphics{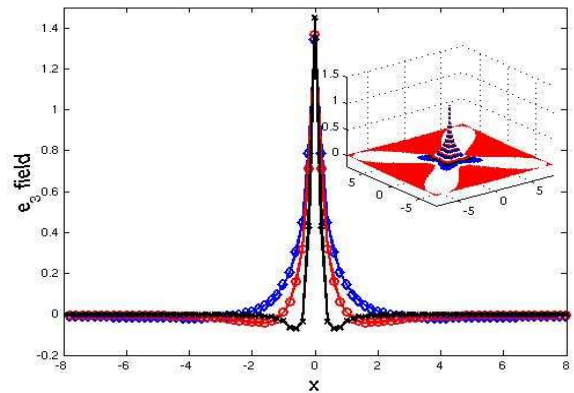}}
\caption{Profile of the strain field $e_3$ along $(Ox)$ for
  frequencies $\omega_0=1$ (blue diamonds), $10$ (red circles), 
  $100$ (black crosses). 
  The smaller $\omega_0$, the wider the profile. The parameters used 
  are $A_1=4$, $A=3$, $g=3$, $\sigma=0.01$. Inset shows the surface plot
  of $e_3$. Notice the anisotropy of the field.}
\label{fig:e3profiles}
\end{figure}

\begin{figure}
\psfrag{e3}{\Huge $e_3$}
\psfrag{distance}{\Huge distance}
\resizebox{0.48\textwidth}{.24\textheight}
{\includegraphics{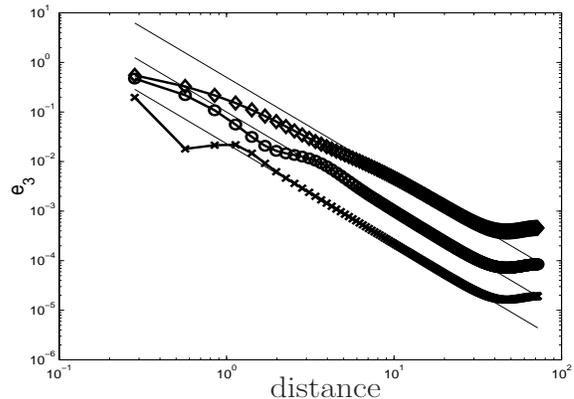}}
\caption{Log-Log plot of the strain field $e_3$ along the diagonal for
  $w_0=1,10,100$ (same symbols as in Fig.~\ref{fig:e3profiles}). We
  have also added $1/r^2$ fits as guides to the eye. The parameters
  used are $A_1=4$, $A=3$, $g=3$, $\sigma=0.01$.}
\label{fig:e3decay}
\end{figure}

These results are illustrated in
Figs.~\ref{fig:e3profiles} and~\ref{fig:e3decay}, showing strain
profiles for different $\omega_0$.
A comment on the underdamped case is in order: an underdamped dynamics
would not remove the discontinuity at $\vec{k}=0$ created by the
compatibility equation; thus, the $1/r^2$ decay is also valid in this
case. The qualitative effect of increasing $\omega_0$ would not be
modified either, although there would be some quantitative
differences from the overdamped case. Finally, we have focused here on
defects created by a locally oscillating $e_3$ strain; the solutions
for locally oscillating $e_1$, $e_2$ strains, or combinations of the 
three strain components, can also be obtained, and are qualitatively
similar. There is one exception to this statement: the functions
$\phi^{(11)}_{\pm}(\vec{k})$ are continuous around $\vec{k}=0$
and even infinitely differentiable. Their tail in real space is thus
{\it exponential} instead of {\it power law}. This implies that the $e_1$
strain field created by a local $e_1$ defect in an isotropic elastic
medium decays exponentially away from the defect with rate $\rho_0$;
if the defect is oscillating, the exponential decay rate
$\rho(\omega_0)$ grows with $\omega_0$ as $\omega_0^{1/2}$. To our
knowledge, this particular case has not been emphasized in the
literature; it would be  interesting to realize its experimental
signatures.

We now consider two defects oscillating with the same frequency
$\omega_0$, with the goal of studying the interactions between them.
Let the two defects be centered at $\vec{r}=\vec{0}$, with phase
$\varphi_1=0$, and $\vec{r}=\vec{r}_0$, with phase
$\varphi_2=\varphi$. We express the fields as linear combinations of
$\phi^{(ij)}_{\pm}(\vec{r})$ and
$\phi^{(ij)}_{\pm}(\vec{r}-\vec{r}_0)$. As above, the twelve complex
coefficients are found by enforcing the regularized oscillating defects
conditions. For each strain configuration, it is
easy to calculate the elastic energy stored in the system from
Eq.~(\ref{eq:energy}). Averaging this energy over one oscillation
period, we obtain the interaction energy of the two defects
$U(\varphi,\vec{r}_0,\omega_0)$. Similar calculations in the static
case were performed, for instance, by Eshelby in Ref.~\cite{Eshelby58}.
\begin{figure}
\psfrag{phi}{\Huge $\varphi$}
\resizebox{0.48\textwidth}{.24\textheight}
{\includegraphics{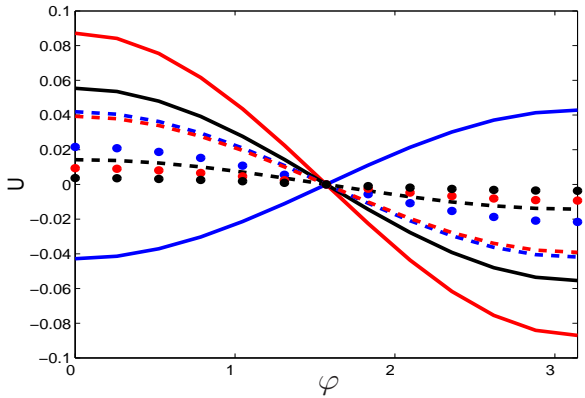}}
\caption{ $U(\varphi$) for two defects along the x-axis. The distance
  between the two defects is $d=2$ (blue curves), $d=4$ (red curves)
  or $d=8$ (black curves). The frequency is $\omega_0=1$ (solid lines),
  $10$ (dashed lines) or $100$ (filled circles). 
The parameters used are $A_1=4$, $A=3$, $g=3$, $\sigma=0.01$.}
\label{fig:energy1}
\end{figure}

\begin{figure}
\psfrag{phi}{\Huge $\varphi$}
\resizebox{0.48\textwidth}{.24\textheight}
{\includegraphics{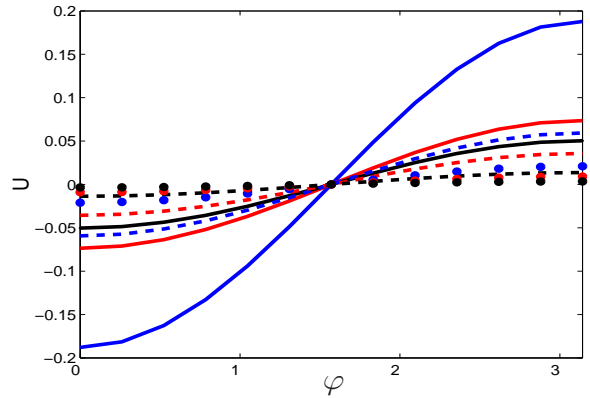}}
\caption{ $U(\varphi$) for two defects along the diagonal. The distance
  between the two defects is $d=1.96$ (blue curves), $d=3.92$ (red
  curves) or $d=7.84$ (black curves). The frequency is $\omega_0=1$
  (solid lines), $10$ (dashed lines) or $100$ (filled circles).  The
  parameters used are $A_1=4$, $A=3$, $g=3$, $\sigma=0.01$.}
\label{fig:energy2}
\end{figure}

We consider first that the two defects are pinned, and study the
function $U(\varphi)$ at fixed $\vec{r}_0$ and $\omega_0$. It turns
out that the energy can be approximated by an XY spin interaction
term,
$U(\varphi,\vec{r}_0,\omega_0)=-J(\vec{r}_0,\omega_0)\cos\varphi$.  A
positive $J$ then corresponds to a ``ferromagnetic'', phase-locking
interaction, and a negative $J$ to an ``antiferromagnetic'' one. As
expected, $J$ decreases to zero at large distances; from the previous
one-defect calculations, it can be anticipated that at fixed distance,
$J$ also approaches zero in the large $\omega_0$ limit. These effects
are seen in Fig.~\ref{fig:energy1}, which shows the energy as a
function of $\varphi$ for two defects along the x-axis.  This figure
also emphasizes a striking effect: $J$ may also change sign with
varying $\vec{r}_0$ or $\omega_0$. For small enough $\omega_0$ and
small enough distance, the interaction is ferromagnetic, and the
defects tend to phase-lock; for larger distances and $\omega_0$, the
interaction is antiferromagnetic, and the energy is minimized for a
maximum phase difference $\varphi=\pi$. Since the fields are
anisotropic, the picture is different for defects situated along the
diagonal; in this case, the interaction is always ferromagnetic, see
Fig.~\ref{fig:energy2}.
   
This qualitative change in the interactions between defects from varying
the frequency or the distance has important consequences when
considering the collective properties of assemblies of defects, as
shown below. We have focused here on oscillating $e_3$
strain fields for specificity; for oscillating $e_1$ or $e_2$ strain
fields, the results are quantitatively different, but the main feature
remains: changes in the frequency or the distance have dramatic
effects on defect interactions.\\

\begin{figure}
\resizebox{0.48\textwidth}{.24\textheight}
{\includegraphics{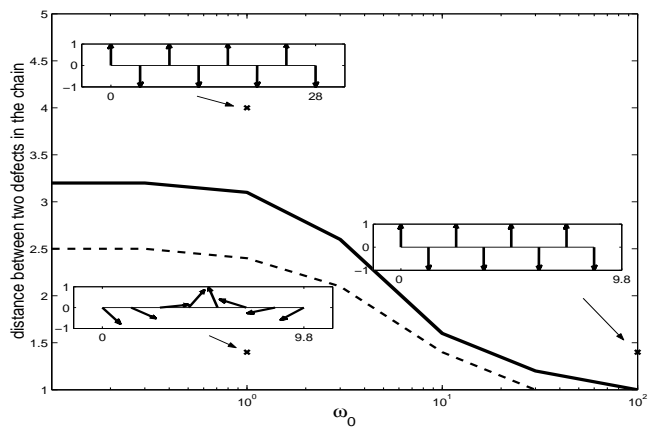}}
\caption{ The solid line is the approximate boundary 
  between ``antiferromagnetic'' and more complex ground states for
  chains of defects parallel to the x or y axis (as predicted by our
  two-defect calculation). The dashed line is the crossover line
  between ``ferromagnetic'' and ``antiferromagnetic'' nearest neighbor
  interactions. The insets show the results of a numerical energy
  minimization for three different cases. The parameters used are
  $A_1=4$, $A=3$, $g=3$, $\sigma=0.01$.}
\label{fig:diag}
\end{figure}

Using a superposition of the elementary solutions $\phi^{(ij)}_{\pm}$,
it is possible to construct the strain profiles associated with an
assembly of $N$ defects. The regularized oscillating conditions have
now to be enforced for each defect, which yields the $6N$ coefficients
of the linear combination. Calculating the energy (averaged over one
period) as a function of all the phases, we are now in a position to
study the collective behavior of the assembly of defects. For the sake
of specificity, we focus again on oscillating $e_3$ strains. As a
working example, we consider chains of equally spaced defects, along
the x-axis or the diagonal, with different spacings and frequencies,
and study the ground state of the system as a function of the phases
of all defects.\\

To obtain a qualitative understanding of the patterning, it is useful
to infer the behavior of $N$ defects from the pairwise interactions
studied above. Note however that summing pairwise
interactions is an approximation, neglecting three-body effects and
beyond. These effects are included in the full numerical calculations. When
defects are aligned along a diagonal, the pairwise interactions are
always phase-locking, see Fig.~\ref{fig:energy2}; there is no
frustration and the ground state should be a perfectly phase-locked,
or ``ferromagnetic'' state, for all lattice spacings and frequencies.
This is indeed observed in our full calculations. The case of defects
aligned along the $x$-axis is richer, see Fig.~\ref{fig:energy1}. As
``antiferromagnetic'' interactions come into play, frustration and
competition between different types of interactions, such as long-period
patterns found in ANNNI spin systems \cite{selke},  are likely to
develop. However, at large enough spacing along the chain and/or large
enough frequency, the interaction between nearest neighbors is
dominant, and ``antiferromagnetic''; the ground state is then expected
to show neighboring defects with a phase difference of $\pi$. At
small lattice spacing and frequency, the interaction between nearest
neighbors is ``ferromagnetic''; the interaction between next nearest
neighbors however is ``antiferromagnetic''. Frustration will then
develop, and a non-trivial ground state may be expected. At
intermediate lattice spacing and frequency, all interactions are
``antiferromagnetic'', but the nearest neighbor interactions do not
dominate; in this case also, a complicated ground state should be
expected. Although this picture based on two-body interactions is
approximate, it provides qualitatively correct results, see
Fig.~\ref{fig:diag}. Figure~\ref{fig:diag} also shows the numerically
determined boundary between ``antiferromagnetic'' ground states (where
nearest neighbor interactions dominate) and more complicated ones, as
expected from the two-defect calculations. The boundary  (dashed line) between
ferromagnetic and antiferromagnetic nearest neighbor interactions is
also sketched. We have performed our calculations here for special
arrangements of defects, but these results demonstrate the existence
of collective behavior, controlled by
the frequency or interparticle distance.\\

In conclusion, we have generalized the well known Eshelby study of static 
elastic inhomogeneities \cite{Eshelby57,Eshelby58} to dynamical oscillating 
defects, focusing on the qualitative properties of the resulting strain 
fields.  Our main results for a single defect are: a $1/r^2$ decay of the
strain fields far from the defect, and that the higher the frequency,
the more localized is the strain perturbation. These results have
important consequences for the interaction between oscillating
defects, which slowly decays at large distances, and is suppressed by
increasing $\omega_0$. A more detailed study of the two oscillating
defects situation shows a more dramatic effect: varying the frequency,
or the distance, can result in a qualitative change in the
interaction, from ``ferromagnetic'' (phase-locking) to
``antiferromagnetic'' (anti-phase-locking). Our work raises the
possibility of 
controlling collective patterning of defects by the frequency or the
defect density; we have explicitly demonstrated such effects by performing
N-defect calculations.\\ 

Applications of our work include non-destructive evaluation of elastic
media with internal oscillating defects, and the collective behavior
of small-polaron dopant sites in directionally bonded electronic
materials~\cite{ref_polarons}. There are natural extensions of our
results to the dynamics of multiple defects. Thus, phase-locking among
defects can be studied if specific phase dynamics controlling the
relaxation pathways is added. Again, if the center-of-mass of the
defects is given dynamics, we can expect self-assembly of defect
patterns driven by the induced elastic fields; since the interaction
range is reduced for higher frequency defect oscillations, this will
reduce the domain of defect ordering. When there are competing ground
states due to ferromagnetic and antiferromagnetic interactions, we
also expect multiscale ``glassy'' dynamics. These extensions of our
results will be reported elsewhere.\\

We are grateful to S.R. Shenoy for insightful discussions.
Work at Los Alamos National Laboratory is supported by the U.S.
Department of Energy.

\end{document}